\documentstyle[11pt,epsf]{article}
\input epsf
\begin{document}
\baselineskip 100pt
\renewcommand{\baselinestretch}{1.5}
\renewcommand{\arraystretch}{0.666666666}
{\large
\parskip.2in
\newcommand{\be}{\begin{equation}}
\newcommand{\ee}{\end{equation}}
\newcommand{\br}{\bar}
\newcommand{\fr}{\frac}
\newcommand{\lm}{\lambda}
\newcommand{\ra}{\rightarrow}
\newcommand{\al}{\alpha}
\newcommand{\bt}{\beta}
\newcommand{\pr}{\partial}
\newcommand{\hs}{\hspace{5mm}}
\newcommand{\up}{\upsilon}
\newcommand{\dg}{\dagger}
\newcommand{\ve}{\varepsilon}
\newcommand{\acc}{\\[3mm]}

\hfill DTP\,98/03

\vskip 1truein
\bigskip
\begin{center}
{\bf Solutions of the Modified Chiral Model in (2+1)
Dimensions.}
\footnote{To Appear in {\bf Journal of Mathematical Physics}}
\end{center}
\vskip 1truein

\bigskip
\begin{center}
T. I{\small OANNIDOU} and W. J. Z{\small AKRZEWSKI}\\
{\sl Department of Mathematical Sciences, University of Durham,\\
Durham DH1 3LE, UK}
\end{center}

\vskip 1truein
\vskip 1truein
\vskip 1truein

{\bf Abstract.}
This paper deals with classical solutions of the modified chiral model on
${\bf R}^{2+1}$.
Such solutions are shown to correspond to products of various factor
which we call time-dependent unitons.
Then the problem of solving the system of second-order partial
differential equations for the chiral field is reduced to solving a
sequence of systems of first-order partial differential equations for the
unitons.
\newpage

{\bf {\rm {\bf I}}. INTRODUCTION}

As is well known \cite{Squires,Ablowitz}, some physical phenomena
(relativity, particle physics, {\it etc}) lead, in their mathematical
description to nonlinear partial differential equations. Some of these
equations are elliptic (when we deal with time independent problems),
others are hyperbolic or parabolic. 
However, with few exceptions, nonlinear partial
differential equations are hard to solve (especially when they involve
more than one spatial variable). Although in some cases static solution
can be found the generalization to the time dependent problems often
presents a formidable task.

A simple model which exemplifies many of the points mentioned above is
the so-called relativistic chiral model in (2+1) dimensions. This model 
involves a chiral field $J(x,y,t)$ which takes its values in $SU(N)$.
The equation of motion for this field is given by
\be
\pr_\mu(J\sp{-1}\,\pr\sp{\mu}J)\,=\,0, \hs \hs J\in SU(N),
\label{chir}
\ee
and is, in fact, the Euler-Lagrange equation corresponding to the
Lagrangian density
\be
{\cal L}\,=\,{1\over 2}\,\mbox{tr}\,\bigl[(J\sp{-1}\pr_\mu
J)(J\sp{-1}\pr\sp\mu
J)\bigr].
\label{lagr}
\ee
Here the indices $\mu, \nu, \al$ range over the values $0,1,2$, with
$x^\mu=(t,x,y)$ and $\pr_\mu$ denotes partial differentiation with respect to
$x^\mu$.
Thus $J$ describes maps of {\bf R$^{2+1}$} into $SU(N)$ and, when we restrict 
our attention to static fields, we have maps from {\bf R$^{2}$} into $SU(N)$.
If we restrict our attention further and consider only static fields whose
energy is finite (see below) then the problem reduces to finding harmonic maps
of ${\bf S}\sp2\rightarrow SU(N)$.

Such harmonic maps were studied by Uhlenbeck \cite{Uhle}
 who, in a seminal paper, 
showed
how to construct these maps in an explicit form. Her construction involved
the introduction of the concept of a uniton. 
She then showed how starting from a given solution one can find a new one
by ``the addition of a uniton".
This observation allowed her to demonstrate that all solutions can be
obtained from the trivial ones ({\it ie} $J$=const) by the addition of a
certain number (in fact $\leq (N-1)$) of unitons.

This beautiful result relied heavily on the integrability of the chiral model
in (2+0) dimensions (for more details see \cite{WJZ}). 
When we consider the same model in (2+1) dimensions,
{\it ie} when we want to solve (\ref{chir}) we immediately run into
a problem as it is not integrable and so Uhlenbeck's method
does not apply.

Some time ago Ward \cite{ward} presented a model in (2+1) dimensions which is
an integrable generalization of the static chiral model in (2+0) dimensions.
The model involves an $SU(N)$ valued chiral field whose equation
of motion now takes the form
\be
(\eta^{\mu \nu}+\varepsilon^{\mu \nu \al}V_\al) \pr_\mu(J^{-1}\pr_\nu J)=0.
\label{mch}
\ee
Here  the tensor $\eta^{\mu \nu}$ is the inverse Minkowski metric,
given by
$\eta^{\mu \nu}=\mbox{diag}(-1,1,1)$; $\varepsilon^{\mu \nu \al}$ is the
fully antisymmetric tensor with $\varepsilon^{012}=1$; and $V_\al$ is 
the constant unit vector pointing in the $x$-direction,  $V_\al=(0,1,0)$.

Notice that although (\ref{mch}) is not relativistically covariant
(as $V_\al$ is nonzero) the solutions of Uhlenbeck are static solutions
of (\ref{mch}). 
As the model is integrable many solutions of (\ref{mch})
can be easily found.  
The first solutions of this model \cite{ward} exhibited 
many properties seen in other integrable models (like wave-like behaviour 
with a phase shift, soliton-like structures passing through each other,  
{\it etc}).
Recently, Ward \cite{warda}, Ioannidou \cite{ioan} and even more recently  
Anand \cite{A} have found further interesting solutions of
(\ref{mch}). 
These solutions described the evolution of soliton-like extended
structures whose properties were quite similar
to properties of solitonic objects in the $O(3)$ ~sigma model ({\it ie}
they exhibit the $90\sp{\circ}$ scattering).

The approach of Ward and Ioannidou involved treating fields of the model
as the limiting cases of the Riemann problem with zeros. In this paper 
we return to Uhlenbeck's construction of static solutions and show that 
her construction can be generalized to give, in general nonstatic, solutions 
of the modified chiral model of Ward ({\it ie} of (\ref{mch})).
Our approach reproduces the solutions reported in \cite{warda} and 
\cite{ioan} and presents an unusual but interesting application 
of Uhlenbeck's approach.

{\bf {\rm {\bf II}}. THE MODIFIED MODEL}

As shown by Ward \cite{ward} the modified chiral model  (\ref{mch})
has the same conserved energy density as the original chiral model
(\ref{chir}) and this energy is given by 
\be
{\cal E}=-\fr{1}{2}\mbox{tr}\left[(J^{-1}J_t)^2+(J^{-1}J_x)^2
+(J^{-1}J_y)^2\right].
\label{energy}
\ee

To specify the problem of finding solutions of  (\ref{mch}) completely we
have to state the boundary conditions satisfied by $J$.
Thus in order to ensure the finiteness of energy of the solutions, we require
\be
J=J_c+J_\theta(\theta)\,r^{-1}+O(r^{-2}), \hs r \rightarrow \infty,
\ee
where $z=r\,e^{i\theta}$ (with $z=x+iy$), $J_c$ is a constant matrix, and
$J_\theta$ is independent of $t$.

Equation (\ref{mch}) is completely integrable, in the sense that can be
written as the compatibility condition for a pair of linear equations
involving a spectral parameter  $\lm$.
Define
\be   
A_z=\fr{1}{2} J^{-1}(\pr_z-\fr{1}{2}  i \pr_t)J, \hs \hs
A_{\bar{z}}=\fr{1}{2} J^{-1}(\pr_{\bar{z}}+\fr{1}{2}  i \pr_t)J,
\label{al}
\ee
where $A_z$, $A_{\bar{z}}$ are $N\times N$ matrices depending on $x,y,t$
but not on $\lm$.

Then the pair of linear operators,
\be
\vartheta_{\bar{z}}+(1-\lm)A_{\bar{z}}, \hs \hs
\vartheta_z+(1-\lm^{-1})A_z,
\ee
with $\vartheta_z=\pr_z+\fr{1}{2}\, i \lm^{-1} \pr_t$,
$\vartheta_{\bar{z}}=\pr_{\bar{z}}-\fr{1}{2}\, i \lm \pr_t$, commute for
all $\lm \!\!\in \!\!{\bf C}^{*}={\bf C}-\{0\}$, as a consequence of
(\ref{mch}).
Therefore one can find a solution of $E_\lm$ of the equations
\begin{eqnarray}
\vartheta_z E_\lm=(1-\lm^{-1})E_\lm A_z, \hs \hs
\vartheta_{\bar{z}} E_\lm=(1-\lm)E_\lm A_{\bar{z}}.
\label{gen}
\end{eqnarray}

\newtheorem{theor}{Theorem}
\begin{theor}
If $J: {\bf R^{2+1}} \rightarrow$ $SU(N)$ is harmonic, then there exist a
$GL(N, {\bf C}$)-valued function $E_\lm$ of $(\ref{gen})$ which is
analytic, holomorphic in $\lm \!\! \in  \!\! {\bf C}^{*}$ and unitary for
$|\lm|=1$ or $\bar{\lm}=\lm^{-1}$  such that $E_{-1}=J$ and
\be
E_\lm=(E_\lm^{-1})^\dg,
\label{aon}
\ee
where $^{\dg}$ denotes complex conjugate transpose.
Conversely, if $E_\lm$ is holomorphic and analytic; and $A_z$,
$A_{\bar{z}}$ defined by $(\ref{al})$ are independent of $\lm$, then
$J=E_{-1}$ is harmonic.
\end{theor}
This follows from \cite{Uhle} and $E_\lm$ is called an extended solution 
corresponding to $J$.
System (\ref{gen}) is overdetermined and in order for a solution $E_\lm$ 
to exist, $A_z$ and $A_{\bar{z}}$ have to satisfy integrability conditions, 
\begin{eqnarray}
(\pr_z+\fr{1}{2}i\pr_t)A_{\bar{z}}+(\pr_{\bar{z}}-\fr{1}{2}i\pr_t)A_z&=&0,
\nonumber
\acc
(\pr_z-\fr{1}{2}i\pr_t)A_{\bar{z}}-(\pr_{\bar{z}}+\fr{1}{2}i\pr_t)A_z
-2[A_{\bar{z}},A_z]&=&0.
\label{condi}
\end{eqnarray}
Notice that, the first equation of (\ref{condi}) is equivalent to 
(\ref{mch}).

{\bf {\rm {\bf III}}. UHLENBECK'S CONSTRUCTION}

It has been known for some time that all static solutions of 
grassmannian models are also solutions of the static chiral model 
\cite{Eichen}; although, not much has been known about other solutions.
However, Uhlenbeck \cite{Uhle} proved that any harmonic map from the 
Riemann sphere ${\bf S}^2$ to the unitarity group $U(N)$ can be factorized
into a  product of a finite number of factors (so-called unitons) 
involving maps into
Grassmannians $G_k({\bf C}^N)$ each of which satisfies a system of
first-order partial differential equations. 
Thus the problem of solving a system of second-order partial 
differential equations for a static harmonic map has been reduced
to having to solve a 
sequence of systems of  first-order partial differential equations for 
the unitons.
This observation can be generalized to the non-static case (\ref{mch}).

We will seek solutions of (\ref{gen})  of the form
\be
E_\lm=\prod_{k=0}^{n}(1+(\lm-1)R_k),
\ee
for some $n$, where $R_k$ are Hermitian projectors of the form 
$R_k=(q_k^\dg\otimes q_k)/|q_k|^2$ ($R_0$ is a constant matrix, 
{\it ie} 0-uniton),  which satisfy some first order differential
equations; and $q_k$ are $N$-dimensional  vectors which, in general,
depend on $z, \bar{z}, t$ and $|q_k|^2=q_k \cdot q_k^\dg$.
Note that a constant $E_\lm$ is a 0-uniton.
Then, following Uhlenbeck, we consider
\be
\tilde{E}_\lm=E_\lm (1+(\lm-1)R_n),
\label{aext}
\ee
and find
\begin{theor}
Let $E_\lm : {\bf C}^{*}\times {\bf R}^{2+1} \rightarrow GL(N,{\bf
C})$ be an extended harmonic map $E_\lm : {\bf R}^{2+1}
\rightarrow SU(N)$ for $|\lm|=1$.
Then $\tilde{E}_\lm=E_\lm (1+(\lm-1)R_n)$ is an extended map for a
Hermitian projection $(R_n)$ if and only if 
\begin{eqnarray}
(1-R_n)(\pr_t R_n-2 i A_{\bar{z}} R_n)&=&0, \nonumber\\
R_n(\pr_{\bar{z}} R_n-A_{\bar{z}} (1-R_n))&=&0,
\label{eq}
\end{eqnarray}
where $A_{\bar{z}}=-(A_z)^\dg$ given by $(\ref{al})$.
\end{theor}

From the definition of the extended solution, 
$\tilde{A}_{\bar{z}}=(1-\lm)^{-1}
\tilde{E}_\lm^{-1}\,\vartheta_{\bar{z}}\tilde{E}_\lm$ and so
\be
\tilde{A}_{\bar{z}}=(1+(\lm^{-1}-1)R_n)\left(A_{\bar{z}}(1+(\lm-1)R_n)
-(\pr_{\bar{z}}-\fr{1}{2}\, i \lm \pr_t) R_n\right),
\label{apo}
\ee
using the fact that $\tilde{E}_\lm^{-1}=(1+(\lm^{-1}-1)R_n) E_\lm^{-1}$
and $A_{\bar{z}}=(1-\lm)^{-1}E_\lm\,\vartheta_{\bar{z}} E_\lm^{-1}$.
But, from Theorem 1 we know that $\tilde{E}_\lm$ is an extended solution
if the $\lm$ and $\lm^{-1}$ terms vanish, which leads to the system
(\ref{eq}).

Next we observe that
\begin{theor}
Suppose that $E_\lm$ is an extended solution and so is $\tilde{E}_\lm=E_\lm
(1+(\lm-1)R_n)$. Then if $A_{\bar{z}}=\fr{1}{2}E_\lm^{-1}
\,\vartheta_{\bar{z}} E_\lm$ and
$\tilde{A}_{\bar{z}}=\fr{1}{2}\tilde{E}^{-1}_\lm\,\vartheta_{\bar{z}}
\tilde{E}_\lm$, they are related by
\be
\tilde{A}_{\bar{z}}=A_{\bar{z}}-\pr_{\bar{z}}R_n+\fr{1}{2}i \pr_t R_n.
\ee
\end{theor}
This follows automatically from the terms in (\ref{apo}) which do not
contain $\lm$. 
In general, by induction, one can show that
\be
A_{\bar{z} n}=(-\pr_{\bar{z}}+\fr{1}{2} i \pr_t) \sum_{k=0}^{n} R_k,   
\ee
and so, that the corresponding $R_k$ satisfy the following equations
\begin{eqnarray}
(1-R_n)\Bigl(\pr_t R_n-2 i \sum_{k=0}^{n-1}(-\pr_{\bar{z}}+\fr{1}{2} i
\pr_t)R_k \,R_n\Bigr) &=& 0,\nonumber\\
R_n \Bigl(\pr_{\bar{z}} R_n -\sum_{k=0}^{n-1}(-\pr_{\bar{z}}+\fr{1}{2}
i\pr_t)R_k  \,(1-R_n)\Bigr)&=&0.
\label{gene}
\end{eqnarray}
Observe that for $n=1$, system (\ref{gene}) gives $\pr_t
R_1 \,R_1=0$ and  $R_1\, \pr_{\bar{z}} R_1=0$; {\it ie} the equations for
the holonomic static solutions (instantons) of the grassmannian models.
This is the case as  
\be
\pr_t R_1\, R_1\equiv q_1\, q_1^\dg\, \fr{ \pr_t q_1^\dg \otimes
q_1}{|q_1|^4}-\pr_t 
q_1^\dg \,q_1\, \fr{q_1^\dg \otimes q_1}{|q_1|^4}=0,
\ee
which is satisfied, if and only if, $q_1$ is independent of $t$.
Using the same argument for $R_1 \pr_{\bar{z}} R_1=0$ 
we deduce that $q_1=q_1(z)$.
For $n \neq 1$, we have more general and, in general, $t$-dependent solutions.

Now, let us concentrate on (\ref{mch}) and discuss the construction
of its solutions.
If we set, $\tilde{E}_{-1}=J$ and $E_{-1}=J_0$ in (\ref{aext}),
then the chiral field $J$ (due to Theorem 2) is of the form 
\be
J=J_0 (1-2 R_n),
\label{pedio}
\ee
and satisfies (\ref{mch}) if $J_0$ does and if $R_n$ satisfies (\ref{eq}).
This describes a time-dependent multi-soliton solution of (\ref{mch})
since it satisfies (\ref{condi}), {\it ie}
\begin{eqnarray}
(\pr_z+\fr{i}{2}\pr_t)\tilde{A}_{\bar{z}}+(\pr_{\bar{z}}-\fr{i}{2}\pr_t)
\tilde{A}_z\!\!\!&\equiv&\!\!\! 
(\pr_z+\fr{i}{2}\pr_t)\left(-(1-2R_n)(\pr_{\bar{z}}
+\fr{i}{2}\pr_t)R_n+(1-2R_n)A_{\bar{z}}(1-2R_n)\right)\nonumber\\
\!\!\!&&\!\!\!+\,(\pr_{\bar{z}}-\fr{i}{2}\pr_t)
\left((1-2R_n)(\pr_z-\fr{i}{2}\pr_t)R_n+(1-2R_n)A_z(1-2R_n)\right)\nonumber\\
\!\!\!&=&\!\!\!(\pr_z+\fr{i}{2}\pr_t)\left(A_{\bar{z}}-(\pr_{\bar{z}}
-\fr{i}{2}\pr_t)R_n\right)\nonumber\\ 
\!\!\!& &\!\!\!+\,(\pr_{\bar{z}}-\fr{i}{2}\pr_t)
\left(A_z+(\pr_z+\fr{i}{2}\pr_t)R_n\right)\nonumber\\
\!\!\!&=&\!\!\! 0,
\end{eqnarray}
using equations (\ref{al}), (\ref{eq}) and  (\ref{pedio}).

So, to construct time-dependent solutions of the $SU(N)$ model (\ref{mch}), 
one can start from a constant solution (zero uniton) and add to it one 
uniton. 
This solution will be nonconstant - but it is static.
Then, to this one uniton, one can add a second uniton, then a third one,
and so on.
As we are going to see, the number of soliton-like structures
in a $n$-uniton
configuration is not related to $n$
({\it i.e.} the procedure of the ``addition of a uniton"
can both increase or decrease the number of unitons). In the static
chiral model (\ref{chir}) Uhlenbeck's theorem shows that for $U(N)$ the
largest uniton number is less than $N$.
In our case, the $t$-dependence of the system (\ref{gene})  
invalidates this argument; the number of unitons can be  arbitrary.
In the static $SU(2)$ case considered by Uhlenbeck the only solutions
are those described by constant matrices (0-unitons) and factors constructed
from holomorphic functions (1-unitons). In the modified chiral model the 0-
and 1-uniton solutions are the same but then, in addition, we have further
solutions corresponding to two and more unitons. These additional
solutions are nonstatic. We do not know, at this stage, whether there is
any bound on the uniton number so that all solutions correspond to field 
configurations of up this number.

{\bf {\rm {\bf IV}}. THE $SU(2)$ AND $SU(3)$ CASES.}

In this section we shall construct and discuss, as an example,  a
2-uniton solution of the $SU(3)$ model (\ref{mch}).
First of all,  observe that all fields of the $SU(2)$ model can be
embedded into the $SU(3)$ model so that the $SU(2)$ solutions are
automatically also solutions of the $SU(3)$ model.
In what follows we will recover some of them as special cases of the
derived expressions.

Let us observe that the following expression describes a non-static
2-uniton solution:
\be
J=K\,\left(1-2\fr{q_1^\dg\otimes q_1}{|q_1|^2}\right)
\left(1-2\fr{q_2^\dg\otimes q_2}{|q_2|^2}\right),
\label{ped}
\ee
where $q_i$ for $i=1,2$ are the three-dimensional  vectors given by
\begin{eqnarray}
q_1&=&(1,\,f,\,g),\nonumber \\
q_2&=&(1+|f|^2+|g|^2)\,(1,\,f,\,g)-2i\,(t
f^{\prime}+h)\,(\bar{f},\,-(1+|g|^2),\,g\bar{f})\nonumber\\
& &-2i\,(tg^{\prime}+h_1)\,(\bar{g},\,\bar{g}f,\,-(1+|f|^2)),
\label{br}
\end{eqnarray}
and $K$ is a constant matrix.
Here  $f$ and $g$ are rational meromorphic functions of $z$.
The field $J$ takes values in $SU(3)$, is smooth everywhere in ${\bf 
R}^{2+1}$ (as the two vectors $q_i$ are nowhere zero). It 
satisfies the boundary condition (\ref{aon}) and the equation of motion 
(\ref{mch}).

Now let indicate how the solution (\ref{ped}) was constructed and how 
further solutions can be obtained.
One way of proceeding is to take the $q_i$  vectors of the form
$q_i=(1,u_i, v_i)$. 
Then, for the 1-uniton solution corresponding to (\ref{gene}), the $q_1$ 
vector $q_1=(1,u_1,v_1)$ (as we have shown) is given in terms of the complex variable $z$. 
Thus  we write $q_1=(1,f,g)$ where $f$ and $g$ are functions of $z$ only.
Then the  system (\ref{gene}) for $n=2$ reduces to the following first 
order differential equations for the function $u_2$ and $v_2$: 
\be
\pr_{\bar{z}}u_2=\fr{G}{(1+|f|^2+|g|^2)}(-u_2+f),\hs \hs
\pr_{\bar{z}}v_2=\fr{G}{(1+|f|^2+|g|^2)}(-v_2+f),
\ee
where $G=\bar{f}^\prime (-u_2+f)+\bar{g}^\prime 
(-v_2+g)+(u_2g-v_2f)(\bar{f}\bar{g}^\prime -\bar{f}^\prime \bar{g})$ and
\begin{eqnarray}
\pr_t u_2&=&2 i \fr{(1+u_2\bar{f}+v_2\bar{g})}{(1+|f|^2+|g|^2)}(u_2(\bar{f}
f^{\prime}+\bar{g} g^{\prime})+\bar{g} (f^{\prime} g-g^{\prime}
f)+f^{\prime}),\nonumber \\
\pr_t v_2&=&2 i
\fr{(1+u_2\bar{f}+v_2\bar{g})}{(1+|f|^2+|g|^2)}(v_2(\bar{f}f^{\prime}+\bar{g}
g^{\prime})-\bar{f} (f^{\prime} g-g^{\prime} f)+g^{\prime}),
\end{eqnarray}
and $^{\prime}$ denotes the derivative with respect to $z$.
A few lines of algebra allow us to show that $q_2$ of (\ref{br}) 
is the general solution of the equations of the above system.
Notice that when $v_i=0$ ({\it ie} $g=0$), (\ref{ped}) 
corresponds to the field $J$ of the $SU(2)$ model (\ref{mch}).

Calculating the energy density for (\ref{ped}) we find,
\begin{eqnarray}
{\cal E}&=&{\cal E}_0+ 4 \mbox{tr} \left((\pr_{\bar{z}}- \fr{i}{2}
\pr_t)R_2\,(\pr_{z}+\fr{i}{2}\pr_t)R_2 \right)
-4\mbox{tr} \left((\pr_{\bar{z}}-\fr{i}{2}\pr_t)R_2\,\pr_z R_2
\,R_2\right)\nonumber\\
& &-4\mbox{tr} \left((\pr_{z}+\fr{i}{2}\pr_t)R_2 \,R_2 \,\pr_{\bar{z}}
R_2\right), 
\end{eqnarray}
where ${\cal E}_0$ is the energy density of the 1-uniton
({\it ie} static) field.
As is well known \cite{Uhle} ${\cal E}_0$ is a total derivative and so, 
probably, also is ${\cal E}$ but, so far, we have not managed to prove this.
However, we have looked at some special cases and have found that the total 
energy $E$, obtained by integrating ${\cal E}$, is quantized in units of
$8\pi$,
{\it ie} is given by $E=8N\pi$. $N$ appears to be related to the number 
of topological structures. 

Let us discuss some special cases:

({\it a}) The embedding of $SU(2)$ solitons, where we set
\be
g=f, \hs \hs \hs \hs h=h_1, 
\ee
Such soliton solutions are localized along the direction of motion; they
are not, however, of constant size. 
Their height (maximum of ${\cal E}$), is time dependent.
In fact, for $(\deg f\geq \deg h)$ no scattering occurs, {\it ie}  
the solitons are located at the centre-of-mass forming a 
totally symmetric ring configuration. 
Otherwise, there are $(\deg f-1)$ static solitons at the centre-of-mass 
of the system accompanied by $N=\deg h-\deg f +1$ solitons (located at  
$t+z^N=0$; since $J$ departs from its asymptotic value $J_c$ when
$tf^\prime+h=0$)
accelerating towards the solitons in the centre, 
scattering at an angle of $\pi/N$, and then decelerating as they separate.
This phenomenon has been observed, firstly, in the $SU(2)$ model (\ref{mch})
(cf. \cite{warda,ioan}) and is a feature of systems (with 
nontrivial topology) admitting topological solitons.

As an example, let us take $f(z)=z$ and $h(z)=z^2$.
The corresponding field represents
a configuration of two solitons that undergo a $90^0$ 
scattering.
The energy density of this solution is 
\be
{\cal
E}=32\,\fr{1+12r^2+12r^4+32r^2t^2+8t^2-16(x^2-y^2)t}{\left[1+12r^4+4r^2
+16(x^2-y^2)t+1+8t^2\right]^2},
\ee
and is symmetric under the interchange $t\ra -t$, $x\ra y$, $y\ra x$. Figure 
1 illustrates what happens close to $t=0$.

\begin{figure}[b]
\unitlength1cm   
\begin{picture}(16,6)
\put(2.5,5){$t=-2$}  
\epsfxsize=13cm
\epsffile{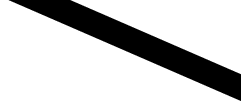}
\end{picture}
\par
\hfill
\begin{picture}(11,6.5)
\put(2.5,4.5){$t=0$} 
\epsfxsize=13cm
\epsffile{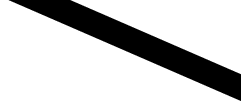}
\end{picture}
\par
\hfill
\begin{picture}(17,5)
\put(2.5,5){$t=2$}   
\epsfxsize=13cm      
\epsffile{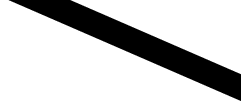}  
\end{picture}
\end{figure}

({\it b}) In a more general case, {\it ie} for arbitrary values of  our 
parameters, the  configuration represents a multi-soliton solution that
is localized close to the centre of the mass and deforms as the time passes.
As an example, let us present a typical case by taking $f(z)=z$, $g(z)=2z$,
$h(z)=z^2$ and $h_1(z)=3$.
In this time-dependent solution, which corresponds to
a configuration of three solitons, the energy density of the field,
for large (negative) $t$, is in the shape of a ring 
which, as time passes, deforms to three peaks and then expands again to a 
ring.
Figure 2 presents pictures of the corresponding energy densities at some
representative values of time.

Returning to the scattering of our soliton-like structures we have seen that
an embedding of the $SU(2)$ model provides an example of such a scattering.
A further example is provided by a genuine $SU(3)$ case when we take
\be
f=\rho^2(z), \hs \hs g=\sqrt{2}\, \rho(z),
\ee
where $\rho(z)$ is an arbitrary function.

One example is presented in Figure 3, where we have chosen  $\rho=z$, $h=z^3$
and $h_1=0$.
This configuration can be seen to consist of two static solitons at the 
centre-of-mass accompanied by two solitons accelerating towards the centre,
scattering at right angles, and then decelerating as they separate.

\begin{figure}[b]
\unitlength1cm
\begin{picture}(5,5.5)
\put(2,5){$t=-35$}
\epsfxsize=13cm
\epsffile{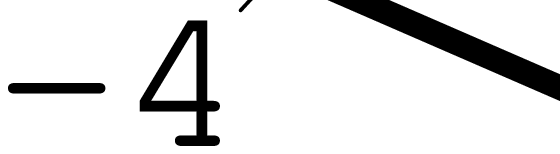}
\end{picture}
\par
\hfill
\begin{picture}(12,6)
\put(2.5,5){$t=-1$}
\epsfxsize=13cm
\epsffile{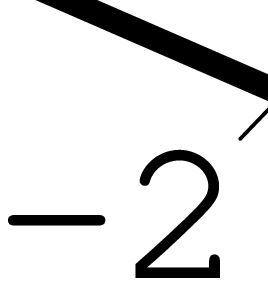}
\end{picture}
\par
\hfill
\begin{picture}(17,5)
\put(2.5,5){$t=0.5$}
\epsfxsize=13cm
\epsffile{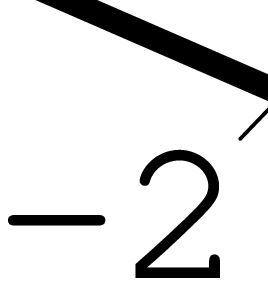}
\end{picture}
\par
\hfill
\begin{picture}(12,7)
\put(2.5,5){$t=30$}
\epsfxsize=13cm
\epsffile{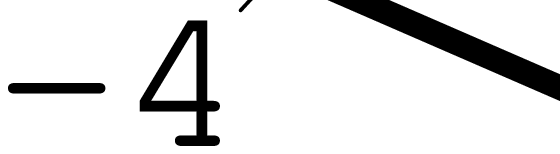}
\end{picture}
\par
\hfill
\end{figure}

\begin{figure}[b]
\unitlength1cm
\begin{picture}(5,5.5)
\put(2.5,5){$t=-15$}
\epsfxsize=13cm
\epsffile{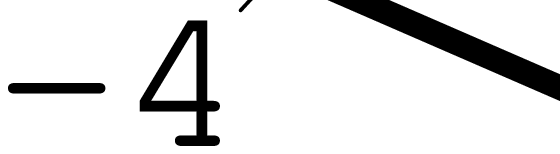}
\end{picture}
\par
\hfill
\begin{picture}(10,5)
\put(2.5,5){$t=-1$}
\epsfxsize=13cm
\epsffile{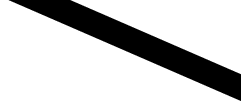}
\end{picture}
\par
\hfill
\begin{picture}(17,3)
\put(2.5,5){$t=0$}
\epsfxsize=13cm
\epsffile{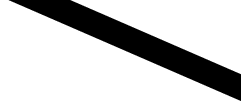}
\end{picture}
\par
\hfill
\begin{picture}(10,5)
\put(2.5,5){$t=1$}
\epsfxsize=13cm
\epsffile{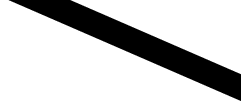}
\end{picture}
\par
\hfill
\begin{picture}(17,3.5)
\put(2.5,5){$t=15$}
\epsfxsize=13cm
\epsffile{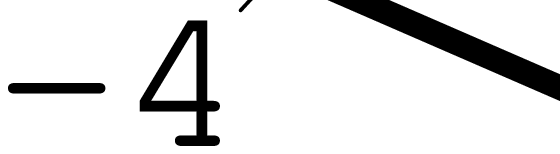}
\end{picture}
\end{figure}

{\bf {\rm {\bf V}}. CONCLUSIONS AND FURTHER COMMENTS}

In this paper we have shown how to adapt Uhlenbeck's construction
of uniton solutions of the chiral model in (2+0) dimensions to the,
time-dependent, uniton-like solutions of the modified chiral model
in (2+1) dimensions. 
It seems likely that there are many more interesting solutions still to be
found. One could, for example, investigate the existence of 
time-dependent solutions based on other static ones constructed by Uhlenbeck.

As for static fields both models are the same, all Uhlenbeck's solutions
are also solutions of the modified chiral model.
However, at the level of two unitons and beyond, our construction
has given us genuine time-dependent solutions of the modified
chiral model. Looking at some examples we have seen that, in some cases,
their field configurations can be thought of as representing scatterings
of some soliton-like objects, sometimes accompanied also by static 
structures.
The total energy appears to be quantized - thus showing that during
their scattering the soliton-like structures must change their size.

We have not succeeded yet in gaining the full understanding of various
properties of our solutions. These problems are currently being 
investigated.

{\bf ACKNOWLEDGMENTS}

We thank R. S. Ward and B. Piette for helpful discussions.
TI acknowledges support from EU ERBFMBICT950035.

{\bf Figure Captions.}

{\bf Figure 1:} Energy density of the  embedding of $SU(2)$ solitons, at
increasing times.

{\bf Figure 2:} Energy density at increasing times, for a three 
ring-shaped solitons.

{\bf Figure 3:} Energy density at increasing times showing a $90^0$
scattering between pure $SU(3)$ solitons.

\end{document}